\documentclass[12pt]{article}
\usepackage{graphics}

\catcode`\@=11 
\def\lsim{\mathrel{\mathpalette\@versim<}}
\def\@versim#1#2{\vcenter{\offinterlineskip
        \ialign{$\m@th#1\hfil##\hfil$\crcr#2\crcr\sim\crcr } }}
\catcode`\@=12 

\begin{document}
\baselineskip=16pt
\begin{titlepage}
\begin{flushright}
TU-702, \ KYUSHU-HET-69, \ hep-ph/0311206
\end{flushright}
\begin{center}
\vspace*{1cm}

{\large\bf 
Radiative CP Phases in Supergravity Theories%
}\vspace{1cm}

Motoi Endo$^{\rm a}$, Masahiro Yamaguchi$^{\rm a}$
and Koichi Yoshioka$^{\rm b}$
\vspace*{5mm}

$^{\rm a}$ Department of Physics, Tohoku University,
Sendai 980-8578, Japan \\
$^{\rm b}$ Department of Physics, Kyushu University,
Fukuoka 812-8581, Japan
\vspace*{2mm}

{\small (November, 2003)}
\end{center}
\vspace*{5mm}

\begin{abstract}\noindent
In this letter, we point out that possible sources of CP violation
originate from radiative corrections to soft terms which are
ubiquitous in supergravity theories and also in other high-energy
frameworks of supersymmetry breaking. With these radiative phases of
gaugino masses and scalar couplings, a complex phase of Higgs
holomorphic mass parameter is generated via renormalization-group
running down to low energy. It is found that its phase value is mainly
controlled by wino as well as gluino, which generally receive
different radiative corrections to their complex phases, even if the
leading part of mass parameters follow from the universality
hypothesis. The radiatively generated phases are constrained by the
existing experimental bounds on electric dipole moments, and may be
detectable in future measurements. They are also found to be available
for the cancellation mechanism to be worked.
\end{abstract}
\end{titlepage}

Low-energy supersymmetry (SUSY) is one of the most attractive
candidates for the fundamental theory beyond the standard model
(SM). It provides various successful applications such as the
stability of mass hierarchy~\cite{hierarchy} and the gauge coupling
unification from the precise electroweak
measurements~\cite{unify}. However supersymmetry must be broken due to
the absence of experimental signatures below the electroweak
scale. Breaking supersymmetry generally gives rise to phenomenological
problems caused by the existence of supersymmetric partners of the SM
fields. One of these problems is the flavor and CP
violation~\cite{CP}. It is usually assumed to overcome the flavor
problem that SUSY-breaking masses of squarks and sleptons are
degenerate within the three generations~\cite{degenerate}. Such a
universality is often discussed in supergravity theory~\cite{GM}. With
this universal assumption, it is clear that the fermion and sfermion
mass matrices are simultaneously diagonalized by superfield rotations
and hence flavor-violating processes are suppressed. It is also
noticed that the universality implies there is no CP phase in
SUSY-breaking scalar masses.

An important point is that CP violation occurs even in the absence of
flavor violation. To see this, we briefly describe conventional
treatment of other four types of parameters in softly-broken
supersymmetric theories. First, gaugino masses are usually assumed to
take a universal value at some high-energy scale. This may be
motivated by the existence of grand unification of the SM gauge
groups. Therefore one has an overall complex phase of gaugino
masses. The renormalization-group evolution (RGE) of gaugino masses
down to low energy does not change their complex phases. Since
scalar trilinear couplings $A$'s carry the flavor indices, the
universal assumption is also adopted for the $A$ parameters to
suppress flavor-changing rare processes. A simply way to realize the
universality is to have vanishing $A$ parameters at high-energy
scale. The RGE of $A$'s is governed by gaugino masses and therefore
generates flavor-blind $A$ terms. Such a scenario may be realized,
e.g.\ by making a separation between SUSY-breaking and visible
sectors. The remaining two parameters are concerned with the Higgs
sector; the supersymmetric Higgs mass $\mu$ and the holomorphic
SUSY-breaking mass $B$. Note that the former suffers from the
so-called $\mu$ problem, that is, how to obtain an 
electroweak-scale $\mu$ parameter. Due to this and related problems,
the situation is rather complicated than the others, and in
particular, the sequestering does not work unlike $A$ parameters (see,
however, dynamical relaxation mechanisms, for example~\cite{B0}) \ We
will simply assume in this letter that $\mu$ is settled to have a
right order of magnitude.

Working with the hypothesis of flavor universality of scalar masses,
we thus obtain four complex parameters in supersymmetric theories; a
universal gaugino mass $M$, a common scalar trilinear 
coupling $A$, supersymmetric Higgs mass $\mu$, and Higgs mixing 
mass $B$. Given that the $U(1)_R$ and Peccei-Quinn rotations can
remove two of these four phases, have we two CP-violating 
parameters $A$ and $B$, where $M$ and $B\mu$ are taken to be real. No
more phases cannot be rotated away by field redefinition. The severest
upper bounds on these two complex phases come from the experimental
results such as non-observation of sizable electric dipole moments
(EDM) of the electron~\cite{eEDM}, neutron~\cite{nEDM} and mercury
atom~\cite{HgEDM}
\begin{eqnarray}
  d_e < 4.3\times 10^{-27} \ {\rm e~cm}, \qquad
  d_n < 6.3\times 10^{-26} \ {\rm e~cm}, \qquad
  d_{Hg}^C < 7\times 10^{-27} \ {\rm cm}.
  \label{bound}
\end{eqnarray}
Here the experimental bound on the EDM of the mercury atom has been
translated into that of the chromo-electric dipole 
moment $d_{Hg}^C$~\cite{FOPR}. For example, in the minimal
supersymmetric standard model, $A$ and $B$ are required to 
satisfy $\arg A\lsim 10^{-1}$ and $\arg B\lsim 10^{-2}$ in the basis
where gaugino masses are real, when the SUSY-breaking masses are a few
hundred GeV.

In this letter, we examine CP-violating phenomena in supergravity
theories. In particular, we point out the importance of
radiatively-generated complex phases of SUSY-breaking parameters,
which often arise inevitably in various frameworks of high-energy
supersymmetry breaking.

\medskip

SUSY-breaking parameters $X$ in general consist of two parts;
\begin{equation}
  X = X_0+\delta X.
\end{equation}
The first term in the right-handed side is the leading contribution
which arises from direct coupling to SUSY-breaking dynamics. We take a
simple and conservative assumption that the leading part, 
e.g.\ of $A_i$'s, can generally be non-universal in size but its phase
is universal. The second term $\delta X$ means sub-leading corrections
in the sense that an absolute value of $\delta X$ is suppressed
compared to the leading part. The point is that these two
contributions are likely to have different origins and hence
independent phase values. In fact, this is indeed the case without
additional assumptions and/or specific dynamics of supersymmetry
breaking. After re-phasing out the overall complex phase of the
leading part, we have a non-vanishing amount of total phase of parameter
\begin{equation}
  \arg X = \frac{1}{X_0}\, {\rm Im}\, \delta X
  +O\bigg(\bigg|\frac{\delta X}{X_0}\bigg|^2\bigg),
  \label{X}
\end{equation}
which cannot be rotated out anymore. The correction $\delta X$ gives
only a few effects on mass spectrum at the electroweak scale and therefore
have been neglected before. However, as we will see below, the
radiatively-induced phases are observable in
CP-violating phenomena as the experimental results tightly constrain
complex phases.

Among various SUSY-breaking parameters, we discuss in this letter the
gaugino masses $M_i$ ($i=1,2,3$) in supersymmetric standard
models. Most generally, possible corrections $\delta M_i$ have
different phase factors, which cannot be re-phased out obviously and
may cause large CP violation. A bit restricted form of corrections we
will encounter is that $\delta M_i$ have a universal phase but their
sizes are different to each other. As an example, consider the
SUSY-breaking masses of the form
\begin{equation}
  M_i = M_0 + \frac{c_ig^2}{16\pi^2}F,
\end{equation}
where $g$ is some coupling constant and $F$ parameterizes a typical
size of SUSY breaking. In the right-handed side of the equation, the
second term denotes the sub-leading part compared to the leading
universal part $M_0$. In this case, non-vanishing complex phases
appear as interference of the two parts. It is found from
eq.~(\ref{X}) that the resultant complex phases at SUSY-breaking scale
are given by
\begin{equation}
  \arg M_i \simeq \frac{c_ig^2}{16\pi^2}\,
  {\rm Im} \bigg(\frac{F}{M_0}\bigg).
  \label{phase}
\end{equation}
Thus radiative corrections to SUSY-breaking parameters, if there
exists, generally become origins of CP breaking. The criterion for
obtaining non-vanishing phases is the existence of corrections which
are (i) ubiquitously seen in the theory and (ii) different in size
between the three SM gauginos. If a theory unavoidably receives such
corrections, one is forced to suppose extra assumptions to control
sizable CP violation.

The relative phases of gaugino masses like eq.~(\ref{phase}) are
detectable in the measurements of EDMs~\cite{EDM}. At the electroweak
scale, that can provide upper bounds on CP-violating phases of
SUSY-breaking parameters. Among them, the severest constraint is
imposed on a phase of Higgs mixing parameter $B$. To estimate a phase
value, it is essential to fix the Higgs mixing mass at some cutoff
scale at which the SUSY-breaking parameters are generated, and solve
the RGEs down to the electroweak scale. The RGE for $B$ is given by
\begin{equation}
  \frac{dB}{dt} = \frac{1}{16\pi^2}\Big(
  6y_t^2A_t+6y_b^2A_b+2y_\tau^2A_\tau+6g_2^2M_2+\frac{6}{5}g_1^2M_1\Big),
\end{equation}
where $y_{t,b,\tau}$ are the Yukawa couplings of the top, bottom and
tau, and $g_{1,2}$ the $U(1)_Y$ and $SU(2)_W$ gauge couplings, 
respectively. A low-energy value of $B$ parameter depends on 
SUSY-breaking parameters at an initial high scale. Its dependence is
described by the approximate solution to the RGE
\begin{eqnarray}
  B(t) &\simeq& B(0) +\bigg(\frac{3}{8\pi^2}\frac{y_t^2(t)}{E(t)}
  \int_0^t\!E(u)du\bigg) A_t(0) \nonumber \\
  && \quad +\sum_{i=1,2,3}\bigg(
  \frac{t}{8\pi^2}r_ig_i^2(t) 
  +\frac{3}{8\pi^2}\frac{y_t^2(t)}{E(t)}
  \int_0^t\!\frac{u}{8\pi^2}r'_ig_i^2(u)E(u)du
  \bigg) M_i(0),
\end{eqnarray}
where the effects of small Yukawa couplings have been neglected.
We have assumed no unification assumption of gaugino masses at the
initial scale, which is relevant to the current interest of 
non-universal corrections. The coefficients $r$'s are fixed by the
charges of corresponding fields and given by $r_i=(\frac{3}{5},3,0)$
and $r'_i=(\frac{13}{15},3,\frac{16}{3})$ 
for $U(1)_Y\times SU(2)_W\times SU(3)_C$. The function $E$ is defined
by $E(u)=\prod_{i=1,2,3}[g_i(0)/g_i(u)]^{2r'_i/b_i}$. We can
understand the result of $B$ parameter as follows. The RGE correction
to $B$ at the electroweak scale is mainly controlled by $M_2$, $M_3$
and $A_t$. In the direct contribution from RGE running, the imaginary
parts of $M_2$ and $A_t$ affect the $B$ parameter. On the other hand,
since a low-energy value of $A_t$ is dominated by the strong gauge
dynamics, so is its phase value. Thus the $M_3$ phase comes into play
in the low-energy $B$ parameter. An initial value of $B$ also directly
appears in the fitting formula. Such behaviors are also easily
understood from the RG-invariant relation among $B$, $A_t$ 
and $M_i$~\cite{RGinv}. In Table~\ref{table:fit}, we present a list of 
one-loop numerical coefficients in the fitting formula for the
electroweak scale $B$ parameter and the EDMs against imaginary parts
of SUSY-breaking parameters at the initial scale. Here we assume the
universal hypothesis defined above and take $|M|=m_0=300$ GeV 
and $A=B=0$ as the leading part of parameters at the cutoff scale. It
is interesting that the phase correction to $B$ comes from the gluino
mass as well as the wino. A total amount of corrections is given by
the interference of these two sizable corrections (the photino mass
effect is negligible due to a tiny gauge coupling). For an
illustration, consider the leptonic EDMs. For not a so small value 
of $\tan\beta$, SUSY radiative corrections are dominated by a one-loop
graph in which the chargino and scalar neutrino propagate in the
loop. This is therefore proportional to $M_2\mu$ and its phase is
given by $\arg(M_2B^*)$. The experimental results tell us that this
quantity must be smaller than $10^{-2}$. From Table~\ref{table:fit},
one can see that the EDM measurements provide severe constraints on
supersymmetric standard models. Given the experimental bounds
(\ref{bound}), the $d_{Hg}^C$ constraint tends to be more restrictive
than the others. However note that we use the chiral quark model for
calculating the EDMs, where there are uncertainties due to some model
dependences and QCD corrections to the EDMs. The QCD uncertainties
also exist in the estimation of the mercury EDM\@.

We thus find that the phase of Higgs mixing parameter at an observable
low scale is induced by radiative corrections through the RGE running,
and inevitably appears at that scale. Such a CP-violating phase can be
large enough to be detectable at the measurements of EDMs. It is also
noted that the $A_t$ phase at an initial scale is restricted as at
comparable level as the $B$ parameter.

\medskip

We now discuss several examples where radiative corrections to
SUSY-breaking parameters naturally appear. If supersymmetry is valid
up to high-energy regime, it is extended to include the gravity. The
gravity multiplet then becomes to mediate SUSY breaking to the visible
sector via super-Weyl anomaly, called the anomaly
mediation~\cite{AM}. It is important that the contribution of the
anomaly mediation is always manifest in supergravity
framework. Moreover, its magnitude is given in terms of anomalous
dimensions of corresponding fields and is different to each
other. Such a contribution has been dropped in the gravity mediation
scenarios because of relative loop suppressions compared to direct
contribution from SUSY-breaking dynamics. However CP violation is
enough sensitive to complex phases including sub-leading
contributions, as we noted before.

\begin{figure}
\begin{center}
\scalebox{0.63}{\includegraphics{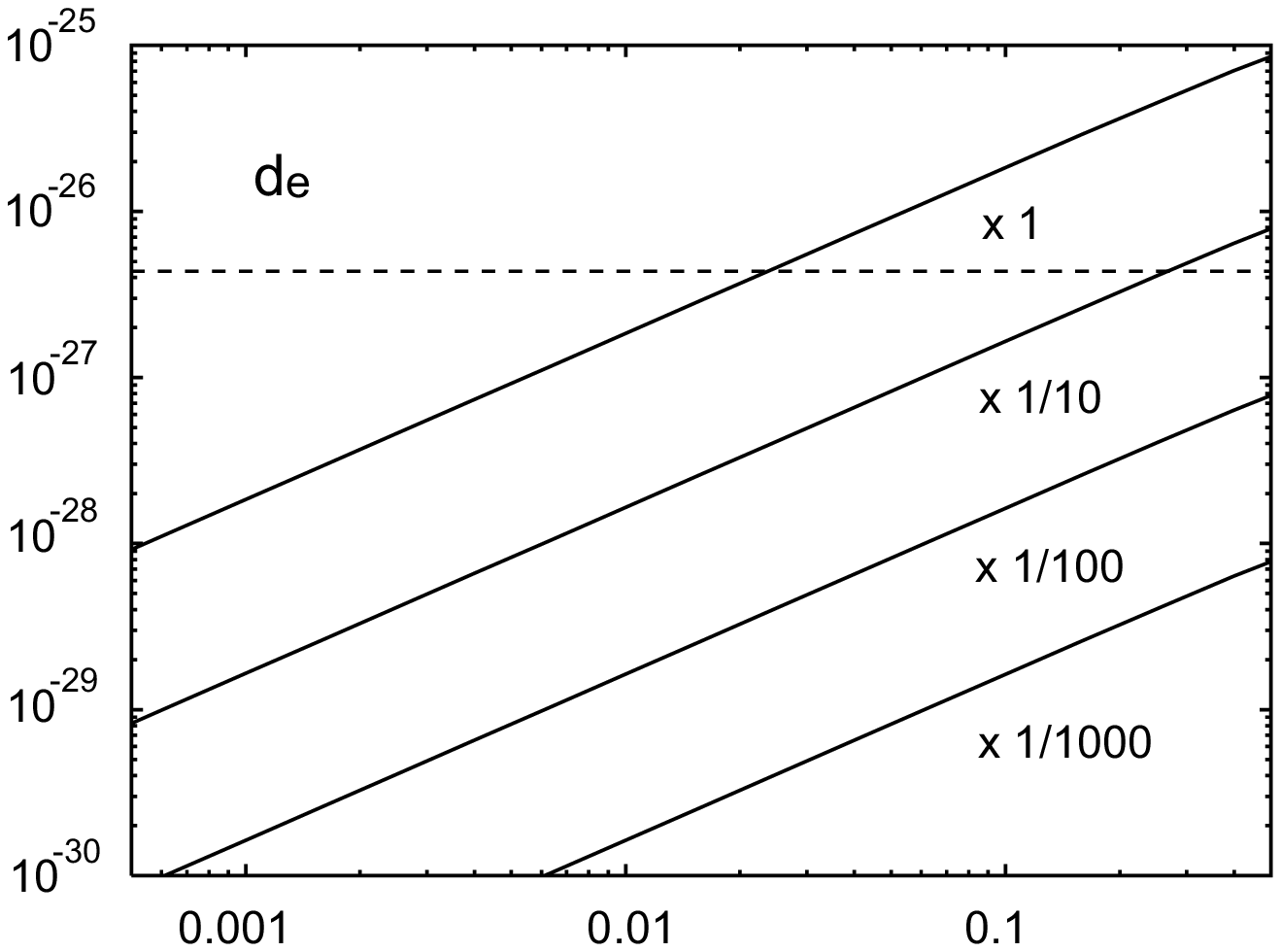}}

\scalebox{0.63}{\includegraphics{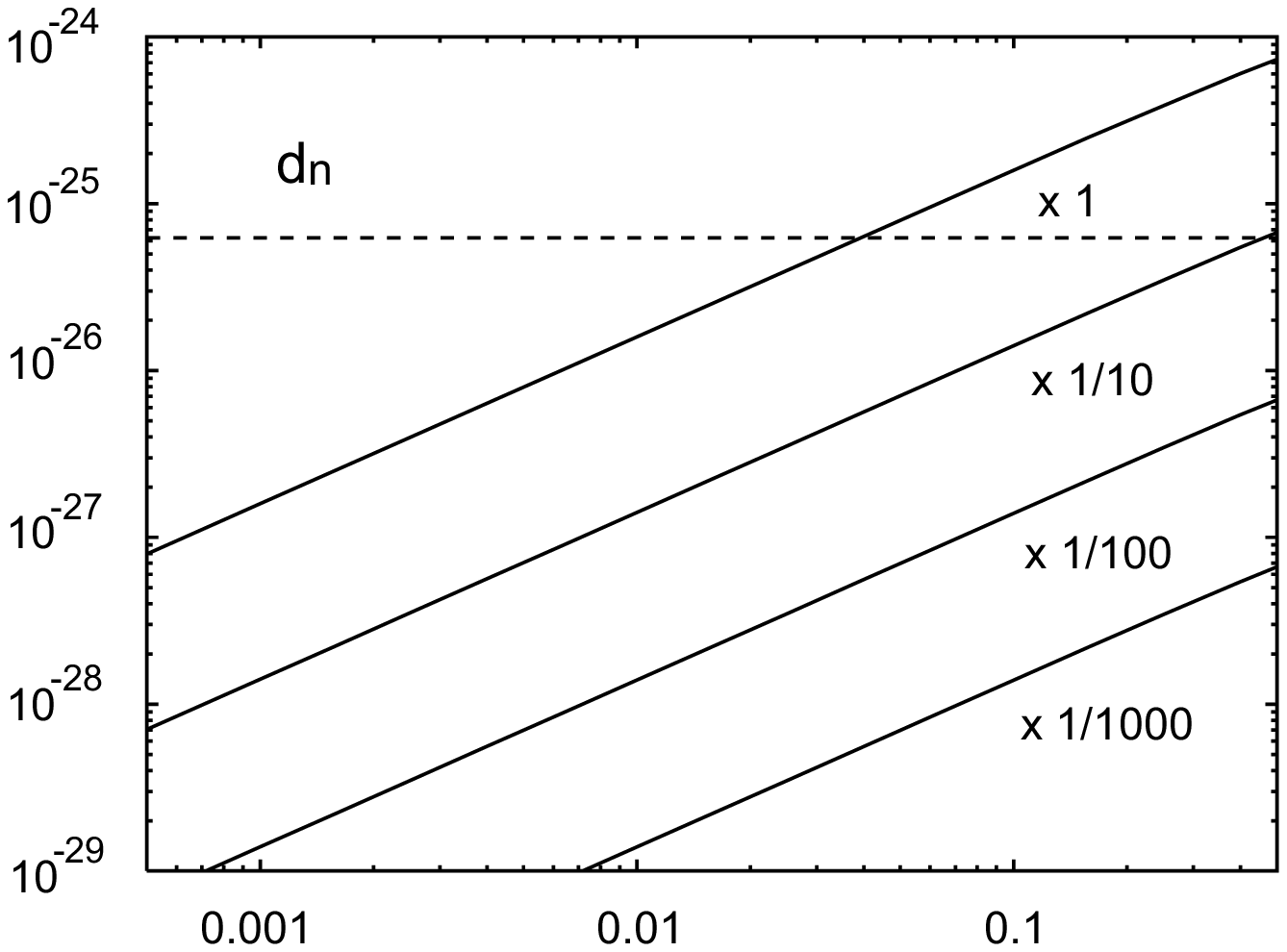}}

\scalebox{0.63}{\includegraphics{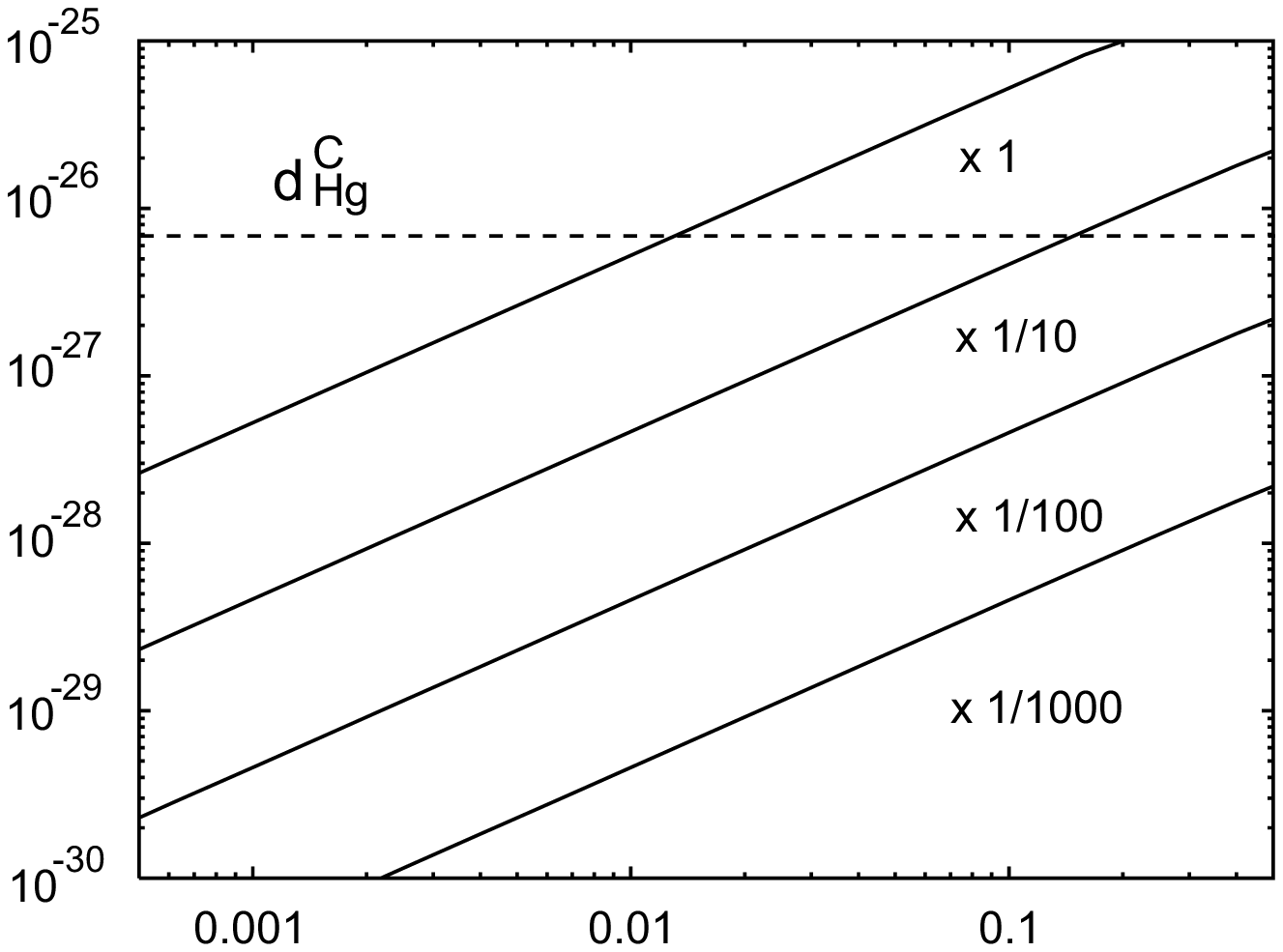}}
\end{center}
\caption{The EDMs in supergravity theory corrected by anomaly-mediated
contribution to gaugino masses. The horizontal axis in each figure is
a relative phase of $F_\phi$ to the leading universal part. The
numbers in the figures denote relative sizes of the corrections. The
initial values of parameters are same as in Table~\ref{table:fit}. 
The dashed lines show the current experimental bounds.}
\label{fig:AM}
\end{figure}

To estimate CP violation, we assume that the leading spectrum follows
from the universality at an initial scale;
\begin{equation}
  M_i(0) = M_0, \qquad A_i(0) = A_0, \qquad B(0) = B_0,
\end{equation}
and the degenerate sfermion masses $m_0$. They come from, e.g.\ the
hidden sector SUSY breaking in supergravity models. In the following
analysis, we take $A_0=B_0=0$, for simplicity. On the other hand, the
ubiquitous radiative corrections appear via the anomaly mediation
whose contributions are
\begin{equation}
  \delta M_i = \frac{\beta_i}{g} F_\phi, \qquad
  \delta A_i = \gamma_i F_\phi, \qquad \delta B = 0,
\end{equation}
where $\beta_i$ and $\gamma_i$ are the gauge beta functions and
anomalous dimensions of matter fields, respectively. Here $\delta B$
is simply assumed to be zero because of unspecified origin of Higgs
mass parameters. This assumption does not change our results unless
some miraculous cancellation occurs among complex quantities. $F_\phi$
is the auxiliary component of the compensator multiplet and gives an
order parameter of SUSY breaking. Requiring a vanishing cosmological
constant, $F_\phi$ is related to other (hidden sector) $F$ terms which
generate the leading part spectrum, then $|F_\phi|\sim |M_0|$. Even if
there is no CP violation in each part of $X$ or $\delta X$, relative
phases generally appears due to the different coefficients 
in $\delta X$'s. Interestingly, the differences of gauge beta
functions are nonzero and model independent as long as preserving the
gauge coupling unification. In Fig.~\ref{fig:AM}, we present a result
of numerical analysis of various EDMs in supergravity scenarios
modified by anomaly mediation. Figures show that the existing
experimental results can detect anomaly-mediated corrections to
gaugino masses, and in turn, put strong restrictions on the sizes and
phases of the corrections. The expected improvements in experimental
precision could give more information about new-physics contribution
such as super-Weyl anomaly and would more severely constrain the model
structure to a non-trivial form.

If the generation of too large complex phases were inevitable,
non-trivial dynamics and/or hypothesis would have to be introduced for
the models to be viable. A naive way is to assume that all parameters
involved in SUSY-breaking dynamics are real. For example, consider the
gravity mediation to induce the leading part of SUSY breaking. In
supergravity, tree-level gaugino masses come from gauge kinetic
functions $f_i=1+\kappa_iZ_i+O(Z^2)$, $Z_i$ denotes a hidden multiplet
responsible for SUSY breaking. At this level, the 
coefficients $\kappa_i$ are required to have a common phase factor,
which can be rotated away by $U(1)_R$ symmetry. However, a combined
analysis with anomaly-mediated corrections means a stronger condition
that $\kappa_i$ must be real without any field redefinition. It is
similarly found that when the leading part is described by the gaugino
mediated contribution~\cite{gauginoM}, a similar condition must be
imposed, that is, one just has to adopt CP-conserving SUSY-breaking
dynamics. On the other hand, the CP phases from (\ref{phase}) allow
two types of possible dynamical resolutions. In the first case, the
phase of leading part is aligned at a high accuracy to the
corrections. One way to realize this situation is the deflected anomaly
mediation scenario~\cite{deflect}. There, SUSY breaking of leading
part is induced by $F_\phi$ effects and the phases are automatically
aligned.\footnote{An alignment mechanism of CP phases will be
discussed elsewhere~\cite{next}, which includes as a simple example
SUSY breaking with the radion stabilization considered in Ref.~\cite{LS}.}
The second is a hierarchy among SUSY-breaking $F$ terms. If the pure
anomaly mediation is the dominant source of SUSY breaking, 
i.e.\ $M_0\ll F_\phi$, CP-violating phases are suppressed. An example
of the inverse type of hierarchy is achieved in gauge mediated SUSY
breaking scenarios~\cite{gauM}. Gauge mediated spectrum is roughly
determined by $F_X/M_X$ where $M_X$ denotes the mass scale of
messenger fields. Therefore the contamination by anomaly mediated
contribution $|F_\phi|\sim |F_X/M_{\rm Pl}|$ is naturally suppressed
for low-scale SUSY breaking $M_X \ll M_{\rm Pl}$. In this case, the
gravitino becomes much lighter than gauginos.

\medskip

Sizable CP-violating corrections could appear in various other
frameworks than the anomaly mediation. It is known that SUSY breaking
in string-inspired supergravity is described in terms of two modulus
fields; the dilaton and the overall modulus. The leading contribution
comes from the dilaton $F$ term which is automatically flavor and CP
blind. On the other hand, (in weakly-coupled theory) the overall
modulus gives one-loop threshold corrections to gaugino
masses. Moreover their sizes depend on gauge beta functions as well as
the Green-Schwarz coefficient. Therefore the criterion to have 
non-vanishing phases is certainly satisfied. As a result, the phases 
of the two modulus $F$ terms must be aligned with some underlying
principle. CP phases from the overall modulus are discussed, 
e.g., in~\cite{KLM}. Another example is grand unified theory
(GUT). Gauge coupling unification is known as one of the motivations
for considering supersymmetry as a promising candidate of new
physics. Then unified gauge group is thought to necessarily break into
the SM group at the GUT scale. This is accompanied by decoupling some
heavy particles, which are the GUT partners of the SM fields. At this
stage, threshold corrections to SUSY-breaking parameters are induced
by these heavy particles circulating in the
loops~\cite{threshold}. It is interesting that these corrections exist 
in any GUT model and give rise to one-loop differences between the 
three gaugino masses, because heavy particle spectrum is GUT breaking 
and split three gaugino masses. As in the case of
anomaly mediation, the corrections generally lead to model-dependent
signatures of EDMs at low energy, which in turn might give an evidence
of grand unification. Radiative phases may also appear at low-energy
thresholds~\cite{threshold2}.

\medskip

\begin{figure}
\begin{center}
\scalebox{1}{\includegraphics{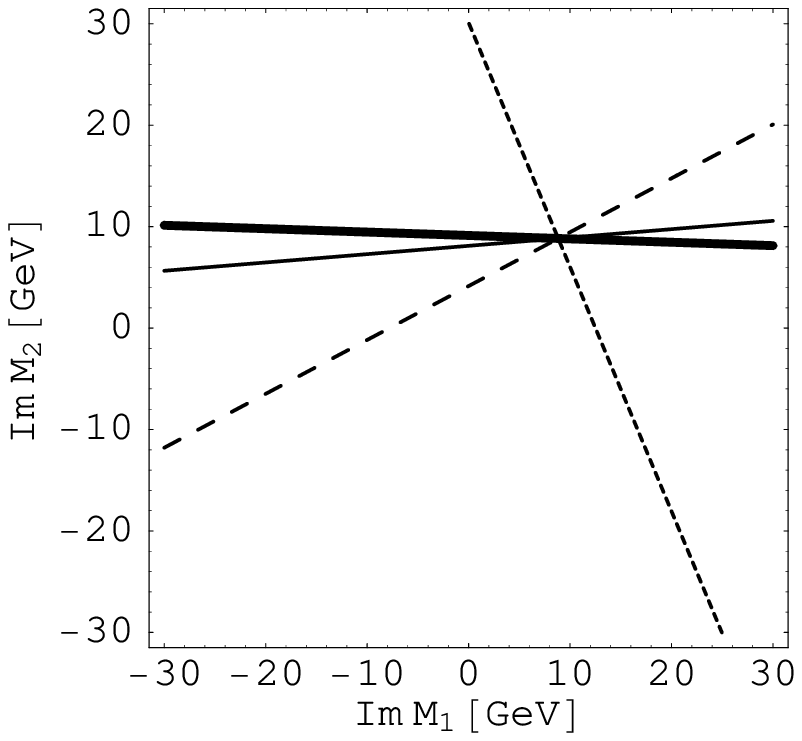}}

\scalebox{1}{\includegraphics{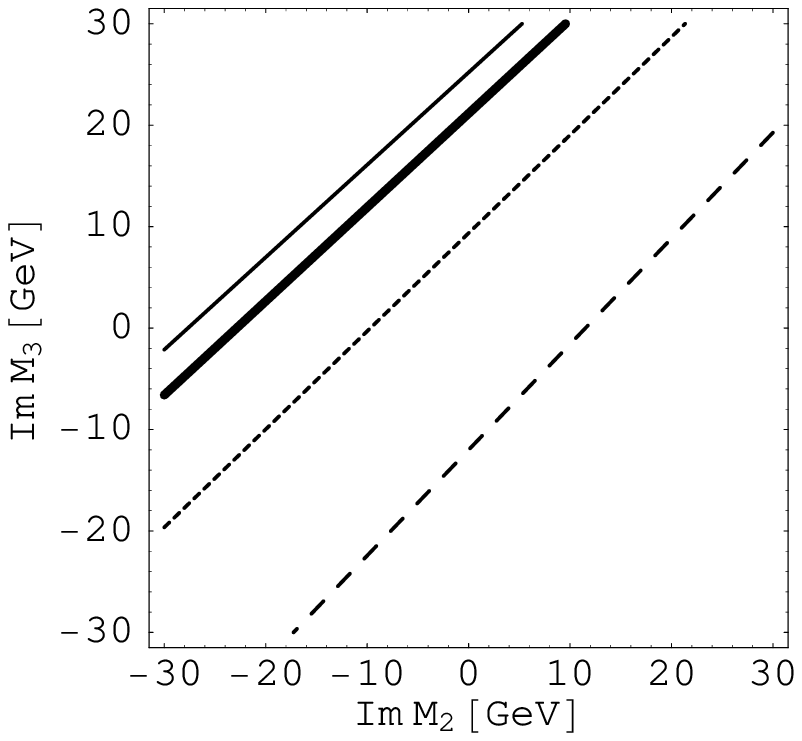}}
\end{center}
\caption{Typical cancellation lines for the EDMs of the electron
(upper graph) and the neutron (lower graph). The bold, solid, dashed
and dotted lines correspond to $N_e=0.1$, 1, 3, 10 and $N_n=1$, 3, 10,
30, respectively. The corrections $\delta A=20i$ and $\delta M_3=20i$
(upper) and $\delta M_1=20i$ (lower) are assumed. The other initial
values of parameters are same as in Table~\ref{table:fit}.}
\label{fig:cancel}
\end{figure}

Finally we mention to another interesting consequence of radiative
phases that they work to ameliorate the CP problem with cancellations
among various diagrams. The cancellation mechanisms with 
possible $O(1)$ phases have been discussed in~\cite{cancel}. There
non-universal spectrum and/or rather large $A$-term contributions are
typically assumed to suppress the EDMs. We now have relative phases of
gaugino masses among the three gauge groups. They are induced
radiatively in a controllable way once high-energy models are
fixed. The phase of the Higgs $B$ parameter is also generated via the
RGE evolution down to the electroweak scale, which phase is described
by those of gauginos. Here we will give a rough estimation of
cancellation of EDMs only in the first-order approximation, and a
complete analysis will be presented elsewhere. First consider the
neutron EDM\@. In the chiral quark model we adopt in this letter, the
neutron EDM is given by $d_n=\frac{4}{3}d_d-\frac{1}{3}d_u$, where the
EDMs of the individual quarks $d_{u,d}$ come from the three
contributions; the electric and chromoelectric dipole moments and the
gluonic dipole moment. The down-quark electric dipole moment gives the
dominant part of the neutron EDM for most parameter space except for
the case of large $\mu$ parameter and small gaugino
masses. Accordingly, severe limits on the CP phases can be avoided 
if $d_d$ vanishes at the electroweak scale, which results in
\begin{equation}
  g_3^2\,{\rm Im}\,(M_3B^*) = g_2^2\,{\rm Im}\,(M_2B^*) 
  N_n(|M_2|,|M_3|,m_Q^2,|\mu|).
  \label{cancel-n}
\end{equation}
We have neglected higher-order terms of the QED gauge coupling 
and $A_d$ term, which is relevant for the case of large $\tan\beta$ 
or $A_d\ll\mu$ (or ${\rm Im}\,(M_3A_d^*)\simeq 0$). The real 
function $N_n$ depends on model parameters and its explicit expression
can be found, e.g.\ in~\cite{cancel}. A similar estimation for the
electron EDM leads to a cancellation condition
\begin{equation}
  g_2^2\,{\rm Im}\,(M_2B^*) = g_1^2\,{\rm Im}\,(M_1B^*) 
  N_e(|M_1|,|M_2|,m_L^2,m_e^2,|\mu|).
  \label{cancel-e}
\end{equation}
The detailed form of $N_e$ is also found in~\cite{cancel}. In
Fig.~\ref{fig:cancel}, we show typical cancellation conditions
(\ref{cancel-n}) and (\ref{cancel-e}) for various values of $N_n$ 
and $N_e$, which depend on model parameters. For an illustration,
we take a single source of radiative corrections, that is, a common
complex phase of the corrections to gaugino masses and $A$
parameters. Even in this restricted case, one can see that the
cancellations do work for wide ranges of parameter space. As an
example, let us consider the corrections from anomaly mediation
discussed before. One first notices that their contribution is
determined by gauge beta functions and leads to a definite model
prediction of induced phases. In Fig.~\ref{fig:cancel}, these
anomaly-mediated corrections are expressed by the lines which are
determined by ratios of gauge beta functions, that are fixed only by
field content of the models. A requirement of CP conservation
therefore could distinguish models. A simultaneous suppression of
various EDMs may be possible for more realistic option with
non-universal radiative corrections. A numerical inspection including
the $d_{Hg}^C$ constraint as well shows that the experimental EDM
constraints actually allow $\arg (M_iB^*)\sim 0.1-0.5$ which are an
order of magnitude larger than naive bounds of phase values. 
The complete analysis rather depends on SUSY-breaking mass spectrum
and we leave it to future investigation, including collider
implications of such large CP phases. Anyway radiative corrections
provide a dynamical justification to adopt the cancellation mechanism
and can make the models to be viable.

\medskip

We pointed out in this letter that 
\begin{itemize}
\item At high-energy scale, gaugino masses and scalar soft terms
receive various radiative corrections in supergravity
theories. It is important that complex phases of these corrections 
can generally differ from the leading part, which phases induce small 
but sizable non-vanishing phases of total soft parameters.
\item The radiatively-induced phases are actually
detectable in EDM measurements via RG evolution of
the phase of Higgs mixing $B$ parameter down to low energy. A 
RG analysis strongly constrains the complex gaugino masses and 
scalar top trilinear coupling at
high-energy scale (Table~\ref{table:fit}).\footnote{After 
submitting the paper, we were aware
that the constraint on the phase of $A_t$ was discussed in
Ref.~\cite{GW}.} These facts give important constraints on models of
SUSY breaking.
\item A cancellation mechanism for suppressing SUSY CP violation can
be worked due to radiative corrections with non-vanishing phases.
\end{itemize}

\medskip

In conclusion, radiative corrections to complex phases of
SUSY-breaking parameters have important consequences for low-energy
phenomenology. Experimental measurements of CP-violating quantities
would select possible model structure through the radiative phase
corrections. It is also possible to cancel out various diagrams of CP
violation as a prediction of the models with controllable phase
parameters.

\bigskip

\subsection*{Acknowledgments}

This work was supported in part by the Grants-in-aid from the Ministry
of Education, Culture, Sports, Science and Technology, Japan,
No.12047201 and No.14046201. ME thanks the Japan Society for the
Promotion of Science for financial support.

\newpage

\begin{table}
\begin{center}
\begin{tabular}{cccccccc} \hline\hline
  & ${\rm Im}\,\delta B$ (GeV) && ${\rm Im}\,\delta A_t$ (GeV)
  && ${\rm Im}\,\delta M_3$ (GeV) && ${\rm Im}\,\delta M_2$ (GeV) \\ \hline
  ${\rm Im}\,B_{\rm EW}$ (GeV) & $1$ && $-0.32$ && $-0.49$ && $0.36$ \\ 
  $d_e$ & $-3.6\times 10^{-26}$ && $1.2\times 10^{-26}$ 
        && $1.8\times 10^{-26}$ && $-1.7\times 10^{-26}$ \\
  $d_n$ & $-3.3\times 10^{-25}$ && $1.1\times 10^{-25}$ 
        && $1.6\times 10^{-25}$ && $-1.5\times 10^{-25}$ \\
  $d_{Hg}^C$ & $-1.2\times 10^{-25}$ && $3.7\times 10^{-26}$ 
                && $4.7\times 10^{-26}$ && $-4.4\times 10^{-26}$ \\ 
  \hline\hline
\end{tabular}
\caption{The fitting formulae for the imaginary part of $B$ parameter
and the various EDMs at the electroweak scale. They depend on the
corrections to SUSY-breaking mass parameters at high-energy scale,
indicated in the first line. For instance, $d_e=-3.6\times 10^{-26}
\times {\rm Im}\,\delta B+1.2\times 10^{-26}\times {\rm Im}\,\delta A_t 
+\ldots$. In the table, we take $|M|=m_0=300$ GeV
and $A=B=0$ as the leading part at the cutoff scale. In our notation,
a scalar trilinear coupling constant is defined as $A\times y$, where
$y$ is a corresponding Yukawa coupling.}
\label{table:fit}
\end{center}
\end{table}

\clearpage

\bigskip

\begin{thebibliography}{99}
\bibitem{hierarchy}
E.~Witten,
Phys.~Lett. B {\bf 105} (1981) 267.

\bibitem{unify}
J.R.~Ellis, S.~Kelley and D.V.~Nanopoulos,
Phys.~Lett. B {\bf 260} (1991) 131;
U.~Amaldi, W.~de Boer and H.~Furstenau,
Phys.~Lett. B {\bf 260} (1991) 447;
P.~Langacker and M.x.~Luo,
Phys.~Rev. D {\bf 44} (1991) 817.

\bibitem{CP}
For earlier work for CP violation in supersymmetric models, 
J.R.~Ellis, S.~Ferrara and D.V.~Nanopoulos,
Phys.~Lett. B {\bf 114} (1982) 231;
W.~Buchmuller and D.~Wyler,
Phys.~Lett. B {\bf 121} (1983) 321;
J.~Polchinski and M.B.~Wise,
Phys.~Lett. B {\bf 125} (1983) 393;
M.~Dugan, B.~Grinstein and L.J.~Hall,
Nucl.~Phys. B {\bf 255} (1985) 413.

\bibitem{degenerate}
S.~Dimopoulos and H.~Georgi,
Nucl.~Phys. B {\bf 193} (1981) 150;
J.R.~Ellis and D.V.~Nanopoulos,
Phys.~Lett. B {\bf 110} (1982) 44.

\bibitem{GM}
A.H.~Chamseddine, R.~Arnowitt and P.~Nath,
Phys.~Rev.~Lett. {\bf 49} (1982) 970;
R.~Barbieri, S.~Ferrara and C.A.~Savoy,
Phys.~Lett. B {\bf 119} (1982) 343;
N.~Ohta,
Prog.~Theor.~Phys. {\bf 70} (1983) 542;
L.J.~Hall, J.~Lykken and S.~Weinberg,
Phys.~Rev. D {\bf 27} (1983) 2359.

\bibitem{B0}
M.~Yamaguchi and K.~Yoshioka,
Phys.~Lett. B {\bf 543} (2002) 189.

\bibitem{eEDM}
E.D.~Commins {\it et. al.},
Phys.~Rev. A {\bf 50} (1994) 2960.

\bibitem{nEDM}
P.G.~Harris {\it et. al.},
Phys.~Rev.~Lett. {\bf 82} (1999) 904.

\bibitem{HgEDM}
M.V.~Romalis, W.C.~Griffith and E.N.~Fortson,
Phys.~Rev.~Lett. {\bf 86} (2001) 2505.

\bibitem{FOPR}
T.~Falk, K.A.~Olive, M.~Pospelov and R.~Roiban,
Nucl.~Phys. B {\bf 560} (1999) 3.

\bibitem{EDM}
For a recent analysis, 
S.~Abel, S.~Khalil and O.~Lebedev,
Nucl.~Phys. B {\bf 606} (2001) 151,
and also references therein.

\bibitem{RGinv}
T.~Kobayashi and K.~Yoshioka,
Phys.~Lett. B {\bf 486} (2000) 223.

\bibitem{AM}
G.F.~Giudice, M.A.~Luty, H.~Murayama and R.~Rattazzi,
JHEP {\bf 9812} (1998) 027;
L.~Randall and R.~Sundrum,
Nucl.~Phys. B {\bf 557} (1999) 79.

\bibitem{gauginoM} 
K.~Inoue, M.~Kawasaki, M.~Yamaguchi and T.~Yanagida,
Phys.~Rev. D {\bf 45} (1992) 328;
D.E.~Kaplan, G.D.~Kribs and M.~Schmaltz,
Phys.~Rev. D {\bf 62} (2000) 035010;
Z.~Chacko, M.A.~Luty, A.E.~Nelson and E.~Ponton,
JHEP {\bf 0001} (2000) 003.

\bibitem{deflect}
A.~Pomarol and R.~Rattazzi,
JHEP {\bf 9905} (1999) 013;
R.~Rattazzi, A.~Strumia and J.D.~Wells,
Nucl.~Phys. B {\bf 576} (2000) 3;
N.~Abe and M.~Endo,
Phys.~Lett. B {\bf 564} (2003) 73.

\bibitem{next}
M.~Endo, M.~Yamaguchi and K.~Yoshioka, work in progress.

\bibitem{LS}
M.A.~Luty and R.~Sundrum,
Phys.~Rev. D {\bf 62} (2000) 035008.

\bibitem{gauM}
M.~Dine, A.E.~Nelson and Y.~Shirman,
Phys.~Rev. D {\bf 51} (1995) 1362;
M.~Dine, A.E.~Nelson, Y.~Nir and Y.~Shirman,
Phys.~Rev. D {\bf 53} (1996) 2658;
For a review, G.F.~Giudice and R.~Rattazzi,
Phys.~Rept. {\bf 322} (1999) 419.


\bibitem{KLM}
S.~Khalil, O.~Lebedev and S.~Morris,
Phys.~Rev. D {\bf 65} (2002) 115014.

\bibitem{threshold}
J.~Hisano, H.~Murayama and T.~Goto,
Phys.~Rev. D {\bf 49} (1994) 1446;
N.~Polonsky and A.~Pomarol,
Phys.~Rev. D {\bf 51} (1995) 6532.

\bibitem{threshold2}
A.~Pilaftsis,
Phys.~Rev. D {\bf 62} (2000) 016007;
Nucl.~Phys. B {\bf 644} (2002) 263.

\bibitem{cancel}
T.~Ibrahim and P.~Nath,
Phys.~Rev. D {\bf 57} (1998) 478;
{\it ibid.} {\bf 58} (1998) 111301;
T.~Falk and K.A.~Olive,
Phys.~Lett. B {\bf 439} (1998) 71;
M.~Brhlik, G.J.~Good and G.L.~Kane,
Phys.~Rev. D {\bf 59} (1999) 115004;
S.~Pokorski, J.~Rosiek and C.A.~Savoy,
Nucl.~Phys. B {\bf 570} (2000) 81.

\bibitem{GW}
R.~Garisto and J.D.~Wells,
Phys.~Rev. D {\bf 55} (1997) 1611.
\end{thebibliography}
\end{document}